# Digital Light Processing in a Hybrid Atomic Force Microscope: *In situ*, Nanoscale Characterization of the Printing Process

*Callie I. Higgins, Tobin E. Brown, *Jason P. Killgore

*Applied Chemicals and Materials Division*

*National Institute of Standards and Technology*

*Boulder, CO 80305*

*corresponding authors: callie.higgins@nist.gov, jason.killgore@nist.gov*

**Abstract**

Stereolithography (SLA) and digital light processing (DLP) are powerful additive manufacturing techniques that address a wide range of applications including regenerative medicine, prototyping, and manufacturing. Unfortunately, these printing processes introduce micrometer-scale anisotropic inhomogeneities due to the resin absorptivity, diffusivity, reaction kinetics, and swelling during the requisite photoexposure. Previously, it has not been possible to characterize high-resolution mechanical heterogeneity as it develops during the printing process. By combining DLP 3D printing with atomic force microscopy in a hybrid instrument, heterogeneity of a single, *in situ* printed voxel is characterized. Here, we describe the instrument and demonstrate three modalities for characterizing voxels during and after printing. Sensing Modality I maps the mechanical properties of just-printed, resin-immersed voxels, providing the framework to study the relationships between voxel sizes, print exposure parameters, and voxel-voxel interactions. Modality II captures the nanometric, *in situ* working curve and is the first demonstration of *in situ* cure depth measurement. Modality III dynamically senses local rheological changes in the resin by monitoring the viscoelastic damping coefficient of the resin during patterning. Overall, this instrument equips researchers with a tool to develop rich insight into resin development, process optimization, and fundamental printing limits.

## 1 Introduction

Three dimensional (3D) printing, or additive manufacturing (AM), is touted as the next generation of agile, efficient manufacturing technology with the ability to fabricate complex structures for applications ranging from inexpensive rapid prototyping to tissue engineering for regenerative medicine.[1–3] Unfortunately, most AM processes introduce micrometer-scale anisotropic inhomogeneities in chemical, thermal, and mechanical properties, causing the performance of fabricated parts to depend strongly and unpredictably on printing conditions.[4,5] Without full understanding of how AM parameters affect mechanical properties, AM will have limited societal impact.[1,6] To fabricate a structure, the desired 3D part file is 'sliced' into discrete 2D slices that are then used to iteratively build the part from a given material. Because AM capitalizes on layer-by-layer fabrication, pure material properties are not necessarily the same as the bulk properties of the final structure due to inherent, uncharacterized heterogeneity introduced within and between each layer.[7]

Focusing on polymers, a variety of AM techniques have been developed to address the needs of specific industries and applications, ranging from nozzle-based systems for biomedical applications to selective laser sintering (SLS) for efficient prototyping. While these techniques produce useful structures, they can be limited in throughput ($\approx$ 1 structure per hour) and in the types of materials used, where the precursor material must be either thermally responsive or thixotropic for nozzle systems and in powder form for SLS.[3,7] Stereolithography (SLA) does not have these limitations and can thus address a wide range of applications including regenerative medicine, prototyping, and manufacturing.

SLA employs a volume of photopolymerizable resin with an illumination source that selectively patterns and polymerizes regions within the material, which allows a wide array of materials to be used. The basic requirements for a resin are that it be photo-reactive, absorbing in the reactive wavelength, and initially in liquid form.[5] This allows AM of materials ranging from hydrogels to reinforced acrylics, with modulus values ranging from $\approx$10 kPa to $\approx$1 GPa with minimum feature sizes that can range from $\approx$1 µm to $\approx$1 mm.[8,9] Furthermore, SLA methods that replace inherently-slow, raster-scan printing with digital light processing (DLP) provide excellent throughput by photo-exposing a full 2D image slice of the desired 3D structure at each layer.[8] During the printing process, patterned light is absorbed by photoinitiator, producing free-radicals that initiate the polymerization. As conversion of monomer to polymer progresses, eventually conversion reaches the gel point, at which time a permanent patterned structure is produced. However, because the process requires light to be absorbed through the thickness of the printed layer in order to print overhanging structures, the mechanical and chemical properties of a printed structure are not currently well defined.[10]

For DLP technology to be harnessed and improved, a fundamental understanding and characterization of AM material properties must be developed. Current techniques used to characterize AM objects, such as tensile and compressive stress testing, do not adequately probe structural heterogeneity because they assume uniform, bulk mechanical properties.[8,11] Critical material properties that impact or result from DLP processing such as cure depth, dose reciprocity, local conversion, and modulus depend on print parameters, cure kinetics, and susceptibility to swelling, none of which have been widely characterized at the individual voxel scale. [5,12–14]Thus, techniques to measure the time-dependent rheological and mechanical properties of photo-patterned structures *in situ* with voxel-scale resolution, are required. However, technologies such as bulk photorheometry, interferometry, and Fourier transform infrared spectroscopy, do not have the combined spatial and temporal resolution to fully capture the dynamic *in situ* environment of DLP.[15–17] The atomic force microscope (AFM) has been extensively developed for applications in polymer science and nanotechnology. Its nanomechanical sensing capabilities are well equipped to measure local variations in the mechanical properties of as-printed DLP parts and photopatterned structures due to the high temporal and spatial resolution of the instrument.[18–20] Furthermore, the proven ability of the AFM to operate in diverse liquid environments affords specific *in situ* characterization potential if applied to the DLP resin.

Probing printed structures during and immediately after fabrication, while still in the liquid resin environment, is critical to fully characterize printed structures because their viscoelastic properties change depending on a complex structure-property-processing relationship.[12,13] Here we present the first-ever hybrid instrument combing a DLP 3D printer with an AFM to characterize the local viscoelastic and swelling behavior of 3D printed voxels throughout the photopatterning process. Recently, we showed how an AFM acting as a local rheometer provides the resolution to probe AM relevant length- (≈100 nm) and time-scales (≈100 μs) during photopolymerization.[13] Here, we use the modular platform of an AFM on an inverted optical microscope as a testbed to deploy a patternable DLP print engine focused at the AFM sensing plane. The system is designed to support a multitude of DLP-relevant modalities, three examples of which are presented here. Modality I demonstrates *in situ* pattern formation and subsequent monomer-swollen nanomechanical characterization to reveal voxel edge heterogeneity and voxel-size dependent mechanical properties. Modality II establishes a novel means of determining the resin gelation point versus part depth and light intensity (i.e. the working curve) *in situ*, with nanoscale resolution. Modality III monitors cure locally throughout polymerization, revealing local, light-intensity-dependent variations in reaction rate, which are critical to characterize the minimum feature size and resolution of the process for a given resin. Combining these modalities offers a cohesive approach to streamline the process of designing and characterizing new resins and to understand the fundamental, underlying print conditions governing structure fidelity, throughput, and performance.

## 2 Materials and Methods

2.1 Formulation of Photopolymer Resins

Materials common to photopolymer additive manufacturing systems were chosen, i.e. acrylate and thiol-ene functional monomers. Pentaerythritol tetrakis(3-mercaptopropionate) (PETMP, 4-arm thiol functional group), tri(ethylene glycol) divinyl ether (TEGDVE , 3-arm -ene functional group), poly(ethylene glycol) diacrylate (PEGDA, acrylate functional group), and diphenyl(2,4,6-trimethylbenzoyl)phosphine oxide (TPO, photoinitiator) were purchased from Sigma-Aldrich‡ and used as received. 2-(2-hydroxyphenyl)-benzotriazole derivative (Tinuvin CarboProtect or TCP, photo-absorber) was purchased from BASF Company and used as received.

Both the thiol-acrylate and thiol-ene resins were prepared within two hours of use. The thiol-ene resin was formulated off-stoichiometry at a molar ratio of 0.64:1 (thiol:ene) to delay the gel point conversion and allow more exposure time in the liquid state. Photoinitiator TPO was included at a final concentration of 10 mg mL$^{-1}$. For the thiol-ene resin, 8 mg TPO was dissolved in 500 µL of tri(ethylene glycol) divinyl ether (TEGDVE) before addition of 300 µL pentaerythritol tetrakis(3-mercaptopropionate) (PETMP), where approximately 10 µL of solution were used per experiment. Conversely, the thiol-acrylate resin was formulated with PETMP and poly(ethylene glycol) diacrylate (PEGDA) off-stoichiometry at a molar ratio of 0.3:1 (thiol:acrylate) to decrease required exposure time to reach gel point conversion. The photoinitiator TPO was included at a final concentration of 29 mg mL$^{-1}$ and the photo-absorber TCP was included at a final concentration of 11.6 mg mL$^{-1}$.

2.2 Hybrid Atomic Force Microscope 3D Printing Instrument

To facilitate *in situ* characterization of the voxel-scale polymerization process, a specialized instrument was developed that combines an atomic force microscope, an inverted optical microscope, and a programmable illumination source. An Asylum Research MFP-3D-BIO was used for the AFM portion of the hybrid-instrument. This AFM utilizes a poly(ether ether ketone) holder design that provides chemical compatibility with many of the resins used in AM. For future needs, the platform also has sample holders that allow for sample heating and user-selected gas environments, both of which are areas of study in photopolymer AM due to the viscosity changes under heating and the inhibiting effects of ambient gases.[21]

Two AFM operational modes were employed in the hybrid instrument: force spectroscopy and Dual AC Resonance Tracking (DART). In force spectroscopy, the force *F* acting on the cantilever tip is monitored while either the cantilever is displaced vertically and bent due to tip-surface interaction, or, when no vertical displacement is present, the surrounding environment is changed in a way that generates a detectable force. When operated with a vertical displacement, the cantilever is brought close to a specimen until a repulsive interaction between the AFM tip and the specimen generates sufficient deflection force as to trigger a retraction of the cantilever (i.e. deflection-triggered force curve). These force curves can be performed over an array of points on the specimen surface, while the force versus indentation response gives information about the modulus of the region sampled. This array method is known as force volume mapping (FVM). The second operational mode, DART, was implemented to track the resonance frequency *f* and quality factor *Q* of the oscillating cantilever during photopolymerization in a recently developed sub-method known as Sample Coupled Resonance Photorheology (SCRPR), where resonance frequency and quality factor are related to the stiffness and viscoelasticity of the sample in question.[22,23] Force modulation (FM) cantilevers (BudgetSensors All-In-One Force Modulation Probe, nominal spring constant of 3.5 N/m) were used for all experiments because the cantilever stiffness and dynamics suitably match the temporal and mechanical property spectra of the AM materials under study.[24]

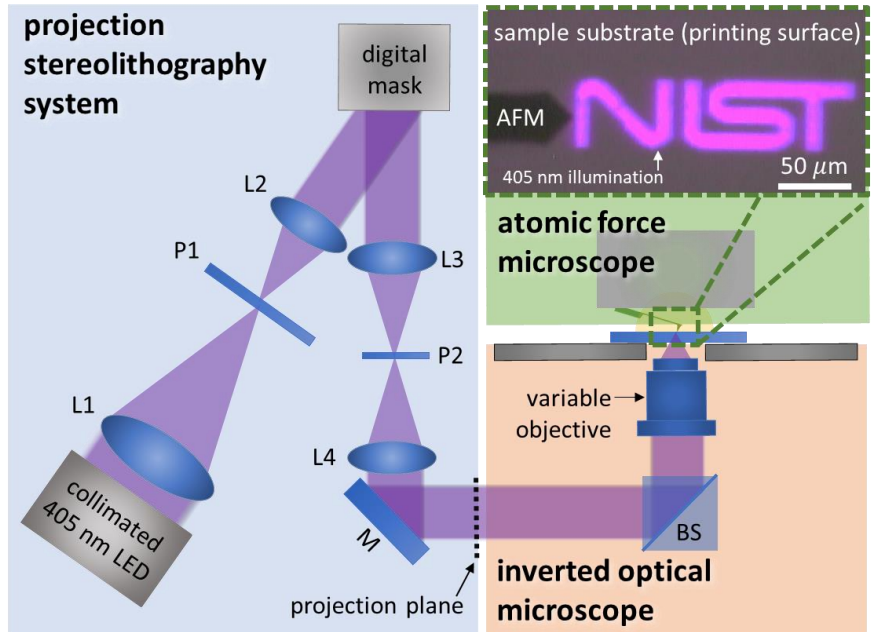

Figure 1: Illustration depicting the hybrid 3D printer AFM. The blue region indicates the digital light processing (DLP) projection system (optical details listed in SI where L = lens, P = polarizer, M = mirror, and BS = beam splitter ), the orange indicates the commercial inverted optical microscope, and the green region indicates the AFM system with an inset optical image of the AFM cantilever and the projected photopattern.

The AFM was mounted onto an Olympus 1X70 inverted optical microscope (Figure 1). This microscope has an external bottom-view illumination port through which we project our photopattern. Exposing the sample from below allows the AFM direct access to the patterning region without blocking the illumination, as would be the case if illuminated from above. The chosen port acts as an external conjugate plane (i.e. a projection in one plane that is directly replicated in another plane with altered dimension) for the focus of any objectives mounted in the inverted microscope.

The ability to change objectives allows for simple changes to the minimum resolution and maximum intensity in the projection system, both of which are important variables to probe in photopolymer AM given their direct influence on part resolution and throughput. For all measurements presented here, a 20X objective (Olympus PLN, numerical aperture *NA* = 0.4) was used. For future measurements, changing objectives from low to high numerical aperture and magnification result in the following relationships that are critical to DLP: decreased depth of field with increased magnification and numerical aperture, increased resolution with increased magnification and numerical aperture, increased intensity with increased magnification, and decreased total projection area with increased magnification. These trade-offs must be considered and optimized prior to choosing the objective for a given experiment.

The photopatterning component is the most novel piece of the hybrid instrument, which employs a DLP photopolymer AM system. This technology was chosen as the representative printing method for this hybrid system because the parallelized nature of DLP presents a significant opportunity for improved throughput in manufacturing applications, thus warranting further investigation. The DLP approach can also approximate a laser raster SLA by increasing its intensity (either at the LED or via magnifying objectives) and by only illuminating single or clustered voxels.

The bulk of the projection system is external to the inverted microscope where a programmable spatial light modulator (SLM) projects a user-defined photomask to the focal plane of the inverted optical microscope/AFM system. The light from a collimated 405 nm light emitting diode ($\lambda$ = 405 nm, M405LP1, Thorlabs) was condensed through a 2-to-1 optical system (including beam reducing optics (Thorlabs) and two z-fold mirrors projection optics (Thorlabs)) to a liquid crystal SLM (E-Series, Meadowlark Optics), which functions as an amplitude mask via two, orthogonally oriented linear polarizers (Edmund Optics) onto which the two-dimensional photopattern is programmed (Figure 1). The patterned light from the SLM is then projected through another 4-F optical system to reimage the pattern at the conjugate plane to the microscope objective's focus, which is located external to the body of the inverted optical microscope. The image is then projected through an objective of choice into the photopolymerizable resin, where either a solid part is formed and characterized by AFM or the AFM is used to dynamically sense the photopolymerization process *in situ*. The light intensity of the hybrid system is calibrated at the focal plane of the objective. This region corresponds to the AFM sensing and DLP printing plane of the system, which allows the measurement to be readily applied to a traditional DLP printer using equivalent exposure conditions. The intensity was measured using an optical power meter (Newport, 2939-R) with a 405 nm calibrated optical power detector (Newport, 918D-SL-OD1R).

The projection is aligned transversely and axially with the sensing portion of the AFM by a series of steps. First, the focus of the projection and the AFM sensing location is obtained by focusing the microscope objective to the top surface of the microscope slide substrate. The projection is then transversely aligned to the AFM sensor (cantilever tip) by monitoring the top-down AFM camera and translating the microscope stage until the illuminated projection is at the desired location with respect to the cantilever tip. After establishing alignment, the projection is turned off and the sample is translated to a region of pristine resin, while maintaining the relative alignment of the cantilever and projection.

At this point, the hybrid system is calibrated to begin experimentation, where the remaining operations are controlled through an external computer. The intensity and exposure time of the photopatterning as well as the AFM sensor location is controlled directly through the AFM user interface. The LED is controlled as an independent input to the AFM controller that is triggered through the AFM software, while the programmed photomask is controlled through the SLM software.

2.3 *In situ* Measurement

The hybrid instrument functions under a variety of *in situ* sensing modalities, three examples of which are presented here. These include a modality to probe the monomer-swollen properties of a voxel after printing, a second modality to probe in situ gelation during photopatterning, and a third modality to sense local rheological changes during the printing process. Notedly, some modalities allow the cantilever to be fully immersed in resin while others presently require that only the cantilever tip be in contact with the resin.

*2.3.1 Modality I: As-printed voxel characterization - Voxel-size Dependence of Mechanical Properties*

The printing environments of photopolymer AM systems are highly dynamic in nature due to diffusion and swelling occurring during the patterning process. Capturing this time-dependent behavior *in situ* is critical to understand, characterize and model the properties of prints due to the associated deformation and property variation throughout the structure. This requires new characterization technologies to probe parts in the native resin-immersed printing environment. Here, the cantilever is fully immersed in a large droplet (≈500 µL) of resin on an acrylate functionalized glass slide, where the acrylate functionality serves to promote covalent bonding of the print layer to the substrate  A deflection-triggered force curve was used to bring the cantilever tip into contact with the underlying glass surface and then retracted a user-defined distance.

Modality I uses the cantilever chip as build-plate against which the desired structure is photopatterned while the entire cantilever chip and cantilever are immersed in resin (Figure 2a). To pattern the sample structure, the photopattern is first aligned directly under the cantilever chip (i.e. the comparatively large millimeter-scale silicon support structure that holds the cantilever), away from the cantilever and tip sensing region. The substrate is then translated to a location with fresh resin and the cantilever is lowered until the tip contacts the substrate, setting a defined layer thickness, $h \approx 60$ µm (dependent on cantilever length), between window substrate and cantilever chip. With cantilever remaining in contact with the window, the photopattern is projected into the resin for a set exposure time and intensity, polymerizing the desired structure against the cantilever chip. Using the cantilever chip as the build plate allows for rapid transition between printing and characterizing. It should be further possible to explore build plate coatings and functionalization. For the current monolithic silicon chip / build-plate, it has the benefit of being low reflection (absorption depth of 0.1 µm at 405 nm), which reduces undesired scattering and enhances print fidelity.[25] Once the exposure is complete, the cantilever is raised to a height above the substrate that is much greater than $h$. This ensures the cantilever tip does not contact the print as the structure is translated to align with the sensing region of the cantilever. Once the sample is aligned with the cantilever tip, the user can obtain a force volume map via force spectroscopy to determine the Young's modulus and topography of the print. To calculate the Young's modulus for all samples, the force-distance curve was fit to a DMT model (Derjaguin, Muller, and Toporov modulus model of solids with adhesion) assuming a tip radius of 20 nm and the Poisson ratio of the sample to be 0.5, with a cantilever spring constant of 3.5 N/m. For the trigger force of 118 nN average tip indentation depth was ≈0.1 μm.

*2.3.2 Modality II: Nanometric Measurement of Gelation and Working Curve*

As with modality I, the cantilever is fully immersed in resin for modality II; however, the photopatterning step takes place at the sensing region of the cantilever, the cantilever tip, and not at the chip (Figure 3a). The photopattern is first aligned with the cantilever tip using the method discussed previously and then the substrate is translated so that patterning occurs at a location with pristine resin. Using a deflection-triggered force curve via force spectroscopy, the cantilever tip is lowered to contact the surface and raised to a user-defined distance above the sample to probe the cure depth dynamics of a given resin at layer height $h$ under two different exposure intensities ($I_0$= 14.5 mW cm$^{-2}$ and $I_0$= 8.7 mW cm$^{-2}$). Here we employ this technique to monitor the bending force acting on the cantilever as the contacting resin undergoes photopolymerization. Specifically, the force detection is synchronized with the photoexposure to determine the duration of time before a force is detected, as well as the magnitude of that force. This mode is designed to sense the gelation front during photopatterning at a range of distances from the exposure focal plane by tracking the cantilever deflection force $F$ as a function of exposure time.

*2.3.3 Modality III: Sample Coupled Resonance Photorheology*

In our previous work, the technique called Sample Coupled Resonance Photorheology (SCRPR) harnessed the high temporal bandwidth of the MEMS cantilever to sense the rapid photorheological response of a photopolymer with ≈ 10 μs temporal resolution.[13] The dynamic response of the AFM probe, indicated by quality factor $Q$ and resonance frequency $f$, was captured throughout the photoexposure. A decrease in $Q$ (in the absence of $f$ change) corresponds to an increase in viscosity while an increase in $f$ corresponds to an increase in stiffness. While valuable, this work was limited to probing a single voxel with limited control of dimensionality and minimum intensity because a laser source was used for exposure. Here, we are now equipped to probe the full DLP cross-section with voxel-level programming of projection intensity.

Because this technique relies on the dynamic response of the cantilever as it oscillates at resonance, it requires that the cantilever not be overdamped by the surrounding fluid. Here, this is accomplished by immersing only the cantilever tip in resin, leaving the cantilever body in air (Figure 4a). This was accomplished by depositing 10 μL of resin onto a glass microscope slide. The edge of a second slide was then drawn across the surface to spread the resin into a thin layer with a thickness of ≈ 4 μm. A nano-needle-tipped FM cantilever (NN-HAR-FM60,

NaugaNeedles, USA) was used, and the third resonance mode (≈ 940 kHz) was tracked while the tip was inserted a known distance into the photopolymer resin. A deflection-triggered force curve was used to bring the needle-tipped AFM cantilever into contact with the underlying glass surface through the resin and then retract ≈ 3 µm to leave the needle immersed in 1 µm of resin. Aside from the resin thickness, the experimental photopatterning step is equivalent to the high intensity exposure step in modality II. During the 60 s exposure duration, DART was used to monitor any changes in $Q$ and $f$. Spatial, rheological changes in the resin were probed during exposure to a circular photopattern where intensity varied radially in a Gaussian distribution where the full-width half-max (FWHM) was 35 µm.

## 3 Results and discussion

To illustrate some of the potential applications of the new instrument, and the insight into the printing process that can be gleaned from its results, data from the three sensing modalities are presented here. These modalities can be broken down into two categories: characterization of the just-printed, resin-immersed voxel, and characterization of the resin and voxel during exposure. These methods collectively characterize the spatial distribution of resin properties in three dimensions, at all stages of the printing process, in a representative photopolymer resin environment, with time and spatial resolution better than the printing process itself.

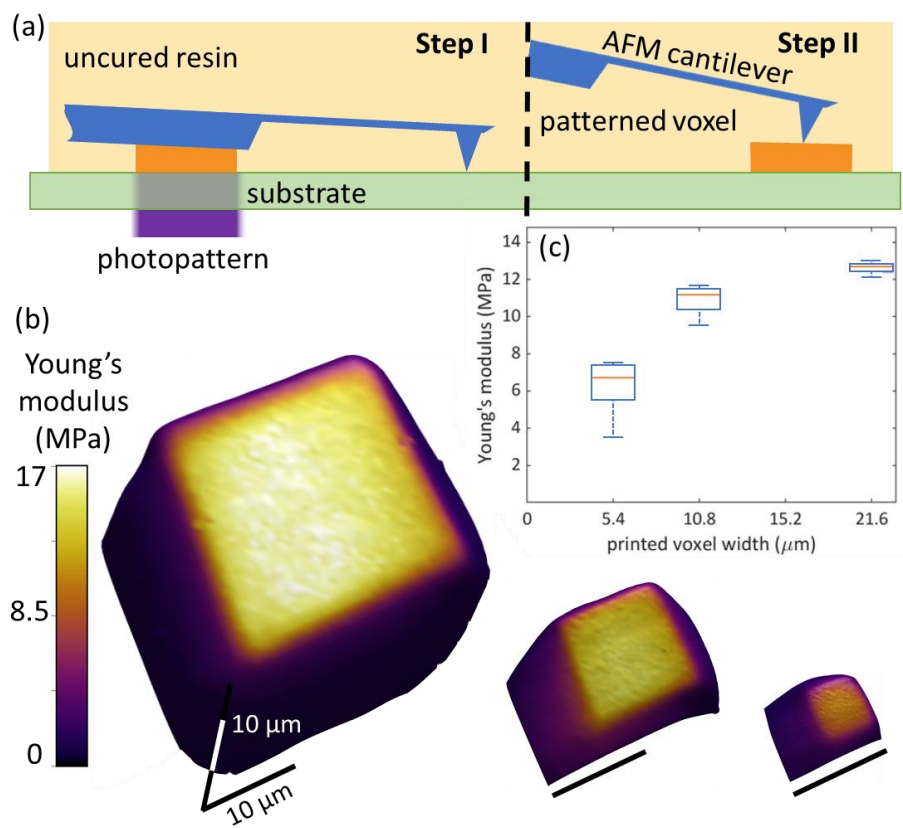

Figure 2: (a) Illustration depicting sensing modality I, where (step I) the voxel is photopatterned against the cantilever chip, which is fully immersed in liquid resin, and then after photopatterning, (step II) the cantilever is translated to a position above the patterned voxel, enabling nanomechanical mapping of modulus variation of the still resin-immersed voxel. (b) 3D rendering of the voxel topography with Young's modulus overlaid in the color channel, the scale bars correspond to 10 µm for all voxels. (c) plots the modulus distribution for each voxel as a function pattern dimension, where modulus systematically increases as pattern dimension increases.

3.1 Modality I: As-printed voxel characterization - Voxel-size Dependence of Mechanical Properties

In photopolymerization processes, a common assumption is that the resultant mechanical properties of a structure will be equivalent for equivalent print energy dose, $E = t_{exp}I_0$, where $t_{exp}$ is exposure time and $I_0$ is light intensity. We recently demonstrated that this reciprocity can break down when intensity and time are varied as the final mechanical properties of a probed photopolymer were not equivalent though $E$ remained equivalent.[13] Another important aspect of reciprocity assumes uniform crosslinking across a photopatterned exposure regardless of voxel size, since the calculation of $E$ does not take into account pattern dimensions. If this assumption was true, the mechanical properties across the solid parts of a single photopatterned layer would be uniform. Modality I tests this assumption by probing the mechanical properties of a structure immediately after printing, as a function of voxel size, while still in the resin environment (Figure 2a).

To probe if mechanical properties are independent of printed feature size for a given set of constant exposure conditions, three different voxel width dimensions $w_v$ were printed, $w_v$ = (21.6 µm, 10.8 µm, 5.4 µm), while keeping exposure time, intensity, and thus dose constant for all voxels ($t_{exp}$ = 20 s, $I_0$ = 14.5 mW cm$^{-2}$, $E$ = 290 mJ cm$^{-2}$) (Figure 2b,c). The resulting moduli are then rendered in 3D with color corresponding to modulus and height to sample topography. We measured the average Young's modulus across each patterned voxel by masking out the measured voxel using the as-programmed voxel dimensions. Typically, the polymerized voxel is larger than the corresponding photopattern, with edges that convolve topographic change and mechanical heterogeneity (Figure 2b). Masking ensures that representative regions are evaluated regardless of final voxel size. We found the $w_v$ = 5.4 µm voxel modulus to be 6.7 MPa and the $w_v$ = 10.8 µm and $w_v$ = 21.6 µm voxels to have average moduli of 11.2 MPa and 12.7 MPa, respectively (Figure 2c). As the average Young's modulus across the three patterned voxels varied significantly, this demonstrates that the assumption of reciprocity is false and likely dependent on the voxel size, illumination conditions, and properties of the resin in question (e.g., diffusivity, reactivity). For example, exposing a circular photopattern with a 100 µm radius into a layer of an acrylate resin with oxygen radical diffusivity on the order of 10 µm$^2$ s$^{-1}$ for 10 s will have inhibitory effects 100 µm into the x-y plane of the photopattern, which is across the entire pattern.

## 3.2 Modality II: Nanometric Measurement of Gelation and Working Curve

The working curve, i.e. the depth of cure versus exposure time, is a critical parameter to photopolymer printing that must be measured for any new resin to inform layer thickness. However, because it is traditionally obtained after the print is post-processed (e.g., washed with solvent, post-cured), it does not adequately capture the swelling behavior of each patterned layer while immersed in resin, which affects all subsequent layers. The hybrid AFM+DLP instrument is able to capture the working curve *in situ* by measuring the cure depth as a function of exposure parameters and distance, $h$, from the patterning substrate (Figure 3). This is done by positioning the cantilever tip directly above the photopattern at different heights $h$ from the substrate and monitoring the swelling force $F$ acting on the cantilever as a function of time, where new locations in the resin were probed for each distance $h$ (Figure 3a).

To illustrate the instrument's ability to probe cure depth, a series of six experiments are presented where the cantilever tip was placed at three defined distances from the substrate, while monitoring $F$ as the photopattern was exposed using two different intensities ($I_0$ = 14.5 mW cm$^{-2}$ and $I_0$ = 8.7 mW cm$^{-2}$) for $t_{exp}$ = 20 s (Figure 3b,c). The three distances represent layer thicknesses $h$ = (1 µm, 10 µm, 28 µm). After initiating illumination, some lag time is observed, followed by an increasing $F$. We hypothesize that the bending force results from swelling of adjacent resin into the just-solidified structure at the instant that resin conversion reaches the gel point. The swelling produces an upward force that is detected by the cantilever (Figure 3b). Thus, the time to detect a specified $F$ at a given $h$ indicates the working curve of the resin. As the tip is moved further from the substrate, gelation is detected at a progressively later exposure duration, and the higher exposure intensity induces a faster response in the polymerizing resin that is positively correlated with exposure intensity, both of which are expected but now experimentally verified (Figure 3b,c).

In addition to the onset of gelation, the rate of swelling within the solid voxel also depends on the tip distance $h$. The rate of swelling, as depicted in the slope of the force versus exposure time curves, is dramatically different for each tip distance. The 1 µm case experienced the slowest rate of swelling and the 28 µm case experienced the most rapid rate of change, regardless of initial exposure intensity. Because most commercial photopatterning AM systems expose layers greater than 28 µm thick, this response is critical to understand and will be the subject of future work. Another notable feature is the decrease in $F$ after $t_{exp}$ = 10 s for the 1 µm 14.5 mW cm$^{-2}$ case, which may be attributed to contributions of gelation and swelling acting on the cantilever body instead of the tip. The final notable observation presented here is the progressive shift in the time at which a given force is detected. The three forces in question, $F$ = (0.005 µN, 0.05 µN, and 0.5 µN), are drawn as horizontal black lines on the graph of Figure 3b,c. The exposure times at which the forces were detected are presented with the specified $F$ in Table 1. At 0.005 µN detection force, the expected dependence of the working curve on $h$ and $I_0$ are recovered. The shortest exposure durations required for detection are observed at high intensity and small $h$, while the longest duration is observed for low intensity and large $h$. Thus, to capture this *in situ* working curve, sensitivity to relatively small (nN) forces is required, which the AFM is well equipped to handle. Further enhanced sensitivity to small forces could be achieved with lower spring constant cantilevers or small cantilevers with reduced thermal noise floor.[14]

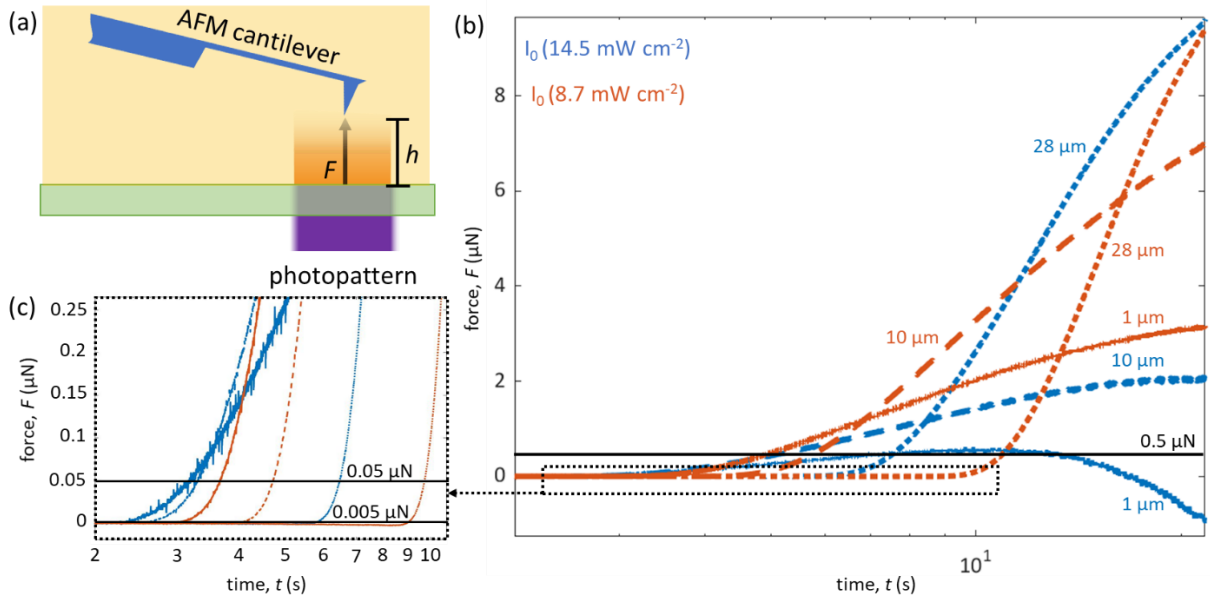

Figure 3: (a) Illustration depicting sensing modality II where the immersed cantilever is positioned some distance *h* directly above the illuminated photopattern region and the force *F* is monitored as a function of exposure time. (b) Cantilever response to the photopolymerization-induced force as a function of exposure time, where the blue and orange lines indicate the two intensities probed ($I_0$ = 14.5 mW cm$^{-2}$ and $I_0$ = 8.7 mW cm$^{-2}$, respectively), and the (-), (--), and (···) lines represent the three distances *h* probed (1 μm, 10 μm, and 28 μm, respectively). (c) A zoomed-in subset of the early exposure time points.

|  |  | Sensor height, *h* | | | | | |
|---|---|---|---|---|---|---|---|
|  |  | 1 μm | | 10 μm | | 28 μm | |
|  |  | *F* (μN) | *t* (s) | *F* (μN) | *t* (s) | *F* (μN) | *t* (s) |
| $I_0$ (mW cm$^{-2}$) | 8.7 | 0.005 | 3.16 | 0.005 | 4.22 | 0.005 | 9.14 |
|  |  | 0.05 | 3.64 | 0.05 | 4.70 | 0.05 | 9.72 |
|  |  | 0.5 | 5.02 | 0.5 | 5.82 | 0.5 | 11.0 |
|  | 14.5 | 0.005 | 2.40 | 0.005 | 2.74 | 0.005 | 5.95 |
|  |  | 0.05 | 3.15 | 0.05 | 3.28 | 0.05 | 6.47 |
|  |  | 0.5 | 7.81 | 0.5 | 5.27 | 0.5 | 7.59 |

Table 1: Table displaying the exposure duration required to reach designated forces $F$ = (0.005 µN, 0.05 µN, and 0.5 µN) for three tip distances $h$ =(1 µm, 10 µm, and 28 µm) and two intensities (8.7 mW cm$^{-1}$ and 14.5 mW cm$^{-1}$). The bold text indicates the tip distance at which the order of duration to gel versus $h$ is inverted.

3.3 Modality III - Sample Coupled Resonance Photorheology

Two questions important for the future of photopolymer AM are: how closely can one pattern adjacent features (resolution) and how small of a feature can one pattern (minimum feature size)? The answers to these two questions are partially dictated by how far reactive species can diffuse and further polymerize, in the cases of crosslinking species, or inhibit, in the case of inhibiting species (e.g. $O_2$), outside of a given photopattern during the projection. The effects of these species are notably present in the photopatterned voxels of Modality I, where the patterned modulus depended monotonically on the programmed voxel size (Figure 2b). Previous SCRPR work also verifies the presence of these relationships to the inverse relationship between the measured viscoelastic damping coefficient ($1/Q$) and probed distance from the exposed voxel.[13] Modality III is designed to investigate this behavior by dynamically probing the rheological response of the resin in response to photoexposure at varying distances from the center of the Gaussian distribution photopattern. Five distances from the center of the circular pattern were probed, $\Delta d$ = (0 µm, 10 µm, 20 µm, 30 µm, 40 µm) (Figure 4b). For this reaction, there was no significant change in resonance frequency $f$, so only $1/Q$ was further considered. The value of $Q$ was captured throughout the photoexposure for all $\Delta d$, where an increase in $1/Q$ corresponds to an increase in viscosity due to polymerization.

For all exposures, $1/Q$ is constant until the photopattern is illuminated, then the cantilever experiences a rapid increase in $1/Q$, indicating the onset of photopolymerization. As $\Delta d$ is increased, the rate and extent of $1/Q$ decreases, as expected for a gaussian photopattern with radial decay in light intensity. In addition, the rapid decrease in the viscoelastic damping coefficient once the exposure is turned off indicates the presence of species diffusion out of the exposure region (Figure 4b,c). This experiment emphasizes the need to calibrate reactive species diffusivity for all resins to optimize and minimize these print dimensions.

Combining sensing modalities II and III allows the effects of gelation (via nanometric working curve measurement) and conversion (via SCRPR) to be separated and sensed in parallel during photopolymerization. By monitoring both the static (i.e. $F$) and dynamic (i.e. $f$ and $Q$) response of the cantilever to the resin on the aligned probe plus pattern system, photopolymerization and swelling behavior can now be observed simultaneously (Figure 4c). The SCRPR mode is sensitive to the onset of polymerization as observed by damping changes at early exposure times, while the static mode is sensitive to local force changes that are most dominant later in polymerization and are not observable using the dynamic mode alone, as highlighted by the arrow in Figure 4c.

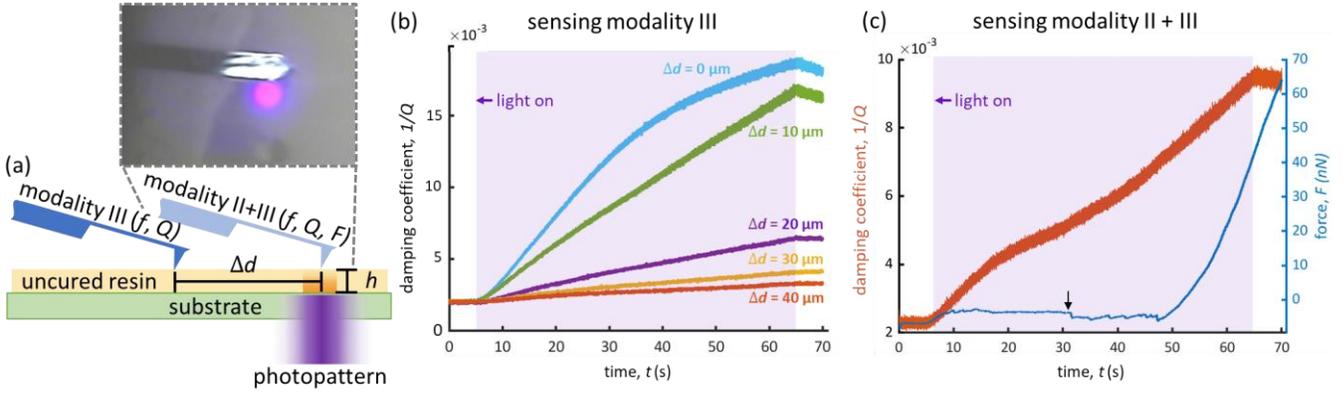

Figure 4: (a) Illustration depicting sensing modalities III and II+III where modality III employs the SCRPR to sense local changes in the viscoelastic properties of the photopatterned resin as a function of distance $\Delta d$ from the center of the photopattern. Inset is a top-down image of the cantilever and photopattern system in the sensing modality III geometry. (b) Graph displaying the detected viscoelastic damping coefficient $1/Q$ as a function of exposure time for probe locations at $\Delta d$ = (0 µm, 10 µm, 20 µm, 30 µm, 40 µm) from the photopattern center. (c) Graph displaying simultaneous sensing in modalities II and III. The increasing damping coefficient $1/Q$ indicates changes in viscosity from polymerization, while the increasing static force $F$ indicates the onset of gelation and swelling.

3.4 Reproducibility

To ensure the reproducibility of each modality, triplicate experiments were conducted and are presented in the Supplementary Information. S2.1 demonstrates Modality I reproducibility by displaying histograms of the Young's modulus for programmed voxels exposed under equivalent conditions ($t_{exp}$ = 20 s, $I_0$ = 14.5 mW cm$^{-2}$) and using equivalent exposure patterns ($w_v$ = 21.6 µm) indicating less than 9 % variation in the mean and similar distribution across the replicate samples. S2.2 highlights Modality II reproducibility via a plot of average force acting on the cantilever during photoexposure for n=3 replicate experiments ($I_0$ =14.5 mW cm$^{-2}$, $h$=28 µm, $t_{exp}$ = 20 s) with one standard deviation shaded or less than 3 % variation between samples. S2.3 similarly demonstrates Modality III reproducibility through a plot of the average of three replicate, equivalent experiments ($I_0$ =14.5 mW cm$^{-2}$, $\Delta d$ = 0 µm, $t_{exp}$ = 60 s) with one standard deviation shaded, showing less than 10 % variation with this measurement.

4    Conclusion

We presented the first-ever hybrid atomic force microscope + digital light processing (AFM + DLP) instrument. While this system potentially supports a multitude of print-relevant measurements, three sensing modalities were presented as exemplary studies. Sensing Modality I probes the just-printed voxel structure and nanomechanics in the printing environment, providing the framework to study the relationships between voxel sizes, print exposure parameters, and voxel-voxel interactions. Modality II captures the nanometric, *in situ* working curve and is the first demonstration of *in situ* cure depth measurement. Modality III dynamically senses rheological changes in the resin by monitoring the viscoelastic damping coefficient of the cantilever during patterning. Reactivity was well correlated with local pattern intensity, indicating utility for local measurement of the reaction profile. Finally, combining Modalities II and III, both the immediate resin conversion and gelation-dependent swelling behavior are captured. This instrument now equips researchers with the tools to develop rich insight into resin development and many other aspects of the photopolymer 3D printing process.

## 5 Acknowledgements and notes